*Katok K.V., Tertykh V.A., Pavlenko A.N., Brichka S.Ya., Prikhod'ko G.P.*

# PYROLYTIC SYNTHESIS OF CARBON NANOSTRUCTURES ON Ni, Co/MCM-41 CATALYSTS

Institute of Surface Chemistry of National Academy of Sciences of Ukraine
03164 Kiev, G. Naumov Str. 17, smpl@ukr.net

*Process of vapor pyrolytic deposition of carbon on nickel and cobalt-containing ordered mesoporous MCM-41 matrices at decomposition of acetylene have been investigated. Formation of nanotubes, nanowires and amorphous carbon particles depending pyrolysis conditions is observed.*

Interest to carbon nanotubes is caused by opportunities of their potential use as scanning probes and sensors, as active elements of field emitters and other electronic devices, as matrice templates for nanocomposites. Recently pyrolytic synthesis of carbon nanostructures from hydrocarbons ($CH_4$, $C_2H_2$, etc.), and also from CO, received general acceptance, and preference is given to the catalytic processes, allowing to obtain not only multilayer, but also single-layer nanotubes with rather high yield.

As a rule, metal catalyst (Ni, Co, Fe) is formed on a surface of a solid by impregnation of matrices by solutions of the corresponding salts of metals with the subsequent formation of catalytically active clusters in the reducing atmosphere. In this regard using of porous matrices based on silica and other inorganic adsorbents are of great interest. In particular, for a synthesis of carbon nanotubes the method of chemical deposition of carbon from a gas phase on porous silica doped with the iron catalyst [1, 2], membranes from the anodized alumina [3, 4], microporous crystals $AlPO_4$ [5] were applied. Because of successes in a synthesis of silicas such as MCM-41 which contain hexagonal structures from uniform cylindrical pores in diameter of 2-10 nanometers, it is expedient to use such mesoporous matrices for obtaining carbon nanostructures. The previous researches have confirmed an opportunity obtaining of nanotubes and carbon nanowires as by graphitization at 800°C different organic templates, which are used for a synthesis of mesoporous silicate structures such as M41S [6], and during pyrolytic deposition of carbon at decomposition of methylene chloride at moderate temperatures (500 and 650°C) [7]. However in the absence of the catalyst on such matrices small yield (up to 2 %) of carbon nanotubes was achieved. At the same time, in work [8] it has been shown that the nanoparticles of nickel supported on mesoporous silica with the purpose of obtaining of conversion catalyst of methane into synthesis gas have stimulated formation of carbon nanotubes on a surface of matrices. Mesoporous silica film doped by iron at a stage of sol-gel synthesis was used as template for obtaining carbon nanostructures at pyrolysis of ethylene at 750°C [2].

The objective of our work is to take the most of advantages of mesoporous matrices such as MCM-41, in particular its uniform structure of pores and developed specific surface, due to chemisorption from a gas phase of the corresponding volatile compounds of the catalyst (acetylacetonates of nickel and a cobalt) and using of moderate conditions of reduction nanoparticles of metal for pyrolytic synthesis of carbon nanostructures (from acetylene).

**Experimental**

Synthesis of ordered mesoporous silica such as MCM-41 was carried out according to recommendations [9] in a reaction mixture with a molar ratio of reagents 1.0TEOS: 0.52CTMABr: $16NH_3$: $39H_2O$ in the following way: hexadecyltrimethylammonium (CTMABr, Aldrich) was introduced into ammonia aqueous solution then tetraethoxysilane (TEOS) was added and mixture before formation gel structure under stirring was heated at 70°C. Obtained sample was calcinated at 540°C for 6 h. Nitrogen ad(de)sorption measurement at 77 K for studying structural characteristics of silica was performed (Micromeritics ASAP-2000 system). Commercial nickel $Ni(acac)_2$ and cobalt $Co(acac)_2$ acetylacetonates (Alfa Aesar) were used as reactants without further purification.

Total scheme of supporting of the catalyst and studies of processes of pyrolysis of acetylene on a surface of silica matrices is presented on Fig. 1.

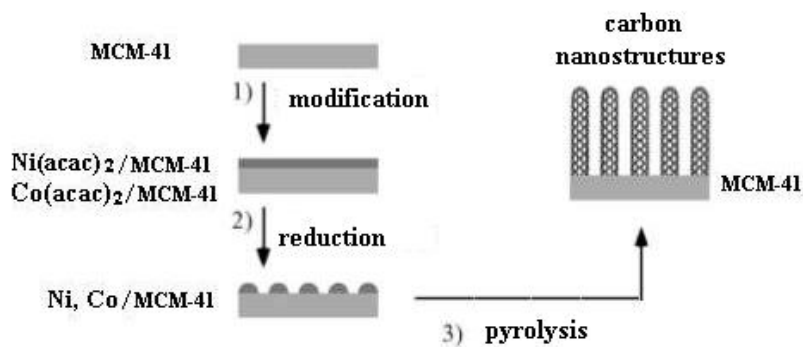

Fig. 1. The scheme of carbon nanostructures obtaining on MCM-41 silica.

*Stage 1 (modification MCM-41 by volatile acetylacetonates of metals).* Modification of mesoporous silica was carried out in a flow-type reactor at atmospheric pressure with use of nitrogen as carrier gas. Procedure of $Ni(acac)_2$ or $Co(acac)_2$ deposition on MCM-41 surface was following: about 0.6 g of powder silica was dehydrated at 250°C for 1 h in air. Then dehydrated sample MCM-41 and

0,1 g of the modifier (acetylacetonate of the corresponding metal) was placed into reactor and was heated up to 150°C. Treatment of silica surface with acetylacetonates of metals was carried out at this temperature for 30 minutes. Subsequently the reactor was cooled in stream of nitrogen to the room temperature.

*Stage 2 (reduction of deposited metal).* A sample of mesoporous silica, modified with acetylacetonate of the corresponding metal, was placed into a horizontal quartz reactor (42 mm inner diameter). Process of reduction of metal in surface compounds was carried out at 450°C for 2.5 h in a flow of hydrogen and argon (1:1 volume ratio). At this stage formation of the large quantity of metal nanoparticles on all surface of support was occurred.

*Stage 3 (pyrolysis of acetylene).* Silica contained nanoparticles of the reduced metal, was used as an active template for the carbon nanostructures growing by pyrolysis $C_2H_2$ at 700°C for 1 h. The scheme of facility is shown on Fig. 2. During pyrolysis of acetylene a deposition of decomposition products of black color on silica matrices is observed.

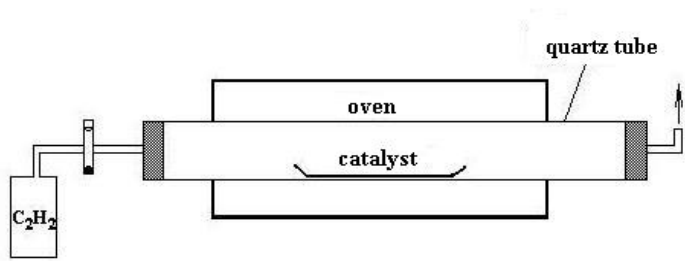

Fig. 2. The scheme of facility for obtaining nanotubes by a method of chemical deposition

The synthesized composite was treated with solution of 44 wt. % HF acid at ambient temperature for removal silica phases. As a result, insoluble carbon fraction was received and used for the subsequent studies.

Identification of carbon structures was carried out using transmission electron microscopy (device JEM-1OOCXII, Japan). The specimens for electron-microscopic studies were prepared by dispersing the particles in alcohol under an ultrasonic treatment and one drop of the suspension was deposited onto a porous carbon film supported on a copper grid.

**Results and their discussion**

Analysis of the isotherm ad(de)sorptions of nitrogen at 77 K on the synthesized silica allows to obtain the information about mesostructure of material. Feature of the isotherm is sharp rise at $p/p_s$=0.25-0.28 and presence of hysteresis loop that corresponds to a capillary condensation in the secondary mesopores at p/p$_s$=0.3-0.9 (fig. 3). Specific surface area of a sample *(S$_{sp}$)*, determined from a linear part of BET equation [10, 11] (at p/p$_s$=0.03-0.2), is equals to 905 m$^2$/g. Total volume of pores *(V$_{tot}$)*, calculated from desorption branches of the isotherm adsorption of nitrogen, using the equation of BET, achieved value of 0.87 sm$^3$/m (at p/ps=0.98). Synthesized sample of MCM-41 is characterized by narrow pores sizes distribution, average diameter of pores, determined by method Barret-Joyner-Halenda [10], is about 39 A.

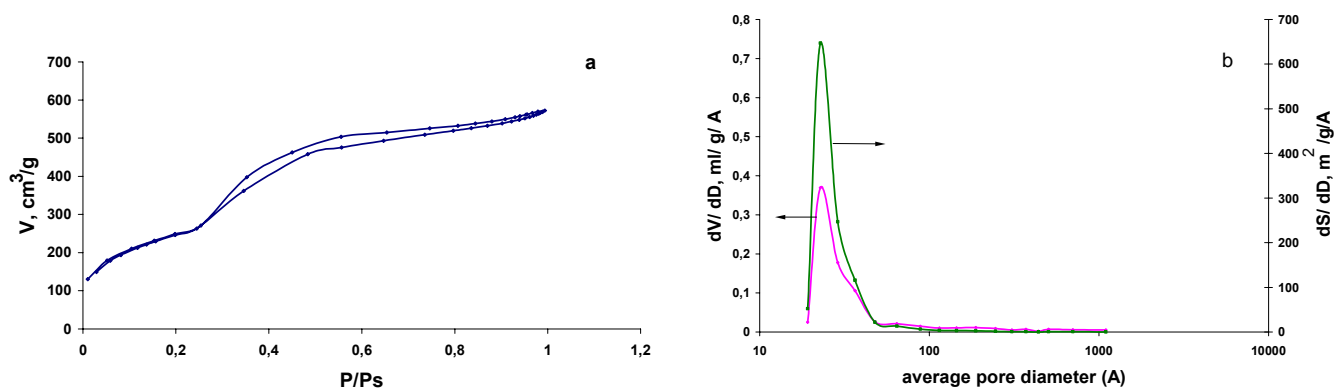

Fig. 3. Isotherm of nitrogen ad(de)sorptions at 77 K (a) and corresponding pore size distribution (b) for silica MCM-41.

Reactivity of acetylacetonates divalent metals in thermodissociation reaction changes in a row: Ni> Cu> Cd> Co [12], that is their thermal stability increases at transferring from a cobalt to nickel. Temperatures of sublimation Ni(acac)$_2$ and Co(acac)$_2$ are accordingly 132 and 97°C. Eliminating of a ligand is observed in the conditions close to temperatures of the beginning of decomposition. All β-diketonates of metals of over-all formula M(acac)$_n$ studied by a method of differential thermal analysis are decomposing in the interval temperatures from 150 up to 400°C [13]. The basic vapor products of decomposition Co(acac)$_2$ are acetone and carbon dioxide, and in the case of Ni(acac)$_2$ - only acetone [12, 14]. To prevent premature decomposition of acetylacetonates of cobalt and nickel, the temperature 150°C was chosen for chemical modification of silica.

As it was established earlier in work [14], under these conditions acetylacetonate of nickel enters into chemical interaction with free silanol groups on surface of silica (absorption band of 3750 cm$^{-1}$ in IR spectrum) with eliminating one ligand. Transferring these observations on other volatile acetylacetonates of metals, we consider that in a case with Co(acac)$_2$ which sublimation temperature is lower, its chemisorption on silica occurs by mechanism of electrophylic substitution of a proton in ≡SiOH groups of a surface by scheme:

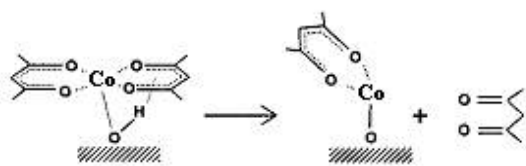

Undoubtedly, such chemical modification of silica by the volatile acetylacetonates of nickel and cobalt provides more uniform supporting of metal in comparison with impregnation of matrices by solutions of the corresponding salts.

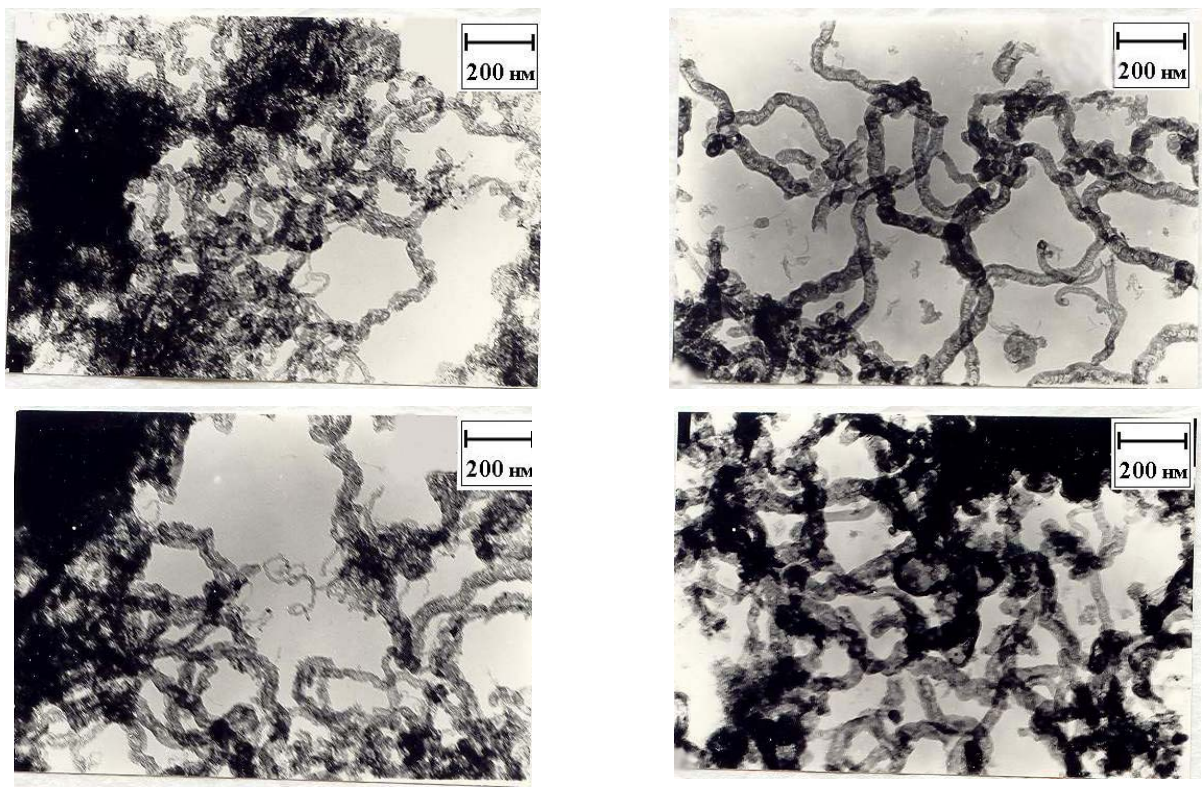

Fig. 4. Micrographs of carbon nanotubes and nanowires, obtained on MCM-41 matrices with nickel (a) and cobalt (b) catalysts.

Using of matrices with the supported metals sufficiently high yield of carbon nanostructures is attained (20-30 % from weight of resulting material). According to electron-microscopic data a lot of nanotubes were formed with external diameters in the range 10-35 nm and length 0,4-0,7 μm for the nickel catalyst (Fig. 4, *a*) and, accordingly 35-52 nm and 0,4-1,5 μm - for the cobalt catalyst (Fig. 4, *b*) are formed. Formation of small quantities of carbon fibers in diameter of 50-70 nm and by length 1,4-2,5 μm is also observed.

Diameter of the obtained carbon nanotubes much more exceeds average pore size of initial silica matrices. Since in these conditions of pyrolysis $C_2H_2$ formation of nanotubes with the sizes close to diameter of pores MCM-41 has not been noted, it is necessary to admit that growth of carbon nanostructures occurs in those places of matrices, where the corresponding nanoparticles of metal are formed (on an external surface of the support, or with possible destruction of walls of hexagonal pores). It is necessary to note also that with increasing in duration of pyrolysis the yield of amorphous carbon grows.

**Conclusions**

The efficient method obtaining of catalysts for a synthesis of carbon nanostructures has been developed by chemisorption volatile acetylacetonates of nickel and cobalt on ordered mesoporous silicas with subsequent reduction of metals in an atmosphere of hydrogen. Depending on conditions of using catalysts, in process of thermodissociation of acetylene the deposition of carbon on silica matrices with formation of carbon nanotubes (diameter of 10-52 nanometers), carbon fibers (diameter of 50-70 nanometers) and amorphous carbon particles is observed.

**Acknowledgements**

The authors are grateful to Yanishpolskii V.V. and Borisenko N.V. for the help in carrying out of experiments.